\documentclass[aps,prl,showpacs,floatfix,twocolumn,amsmath,amssymb,superscriptaddress]{revtex4-1}
\usepackage{graphicx}
\usepackage{color}
\usepackage{dcolumn}

\newcommand{\hay}{$\alpha$-MgCu$_3$(OH)$_6$Cl$_2$}
\newcommand{\hayd}{$\alpha$-MgCu$_3$(OD)$_6$Cl$_2$}

\begin{document}

\title{Haydeeite: a spin-1/2 kagome ferromagnet}
\author{D.~Boldrin}
   \affiliation{University College London, Department of Chemistry,  20 Gordon Street, London, WC1H 0AJ, UK}
\author{B.~F\aa k}
   \affiliation{Institut Laue-Langevin, CS 20156, F-38042 Grenoble Cedex 9, France}
   \affiliation{CEA and Universit\'e Grenoble Alpes, INAC-SPSMS, F-38000 Grenoble, France}
\author{M.~Enderle}
   \affiliation{Institut Laue-Langevin, CS 20156, F-38042 Grenoble Cedex 9, France}
\author{S.~Bieri} 
   \affiliation{Laboratoire de Physique Th\'eorique de la Mati\`ere Condens\'ee, Universit\'e Pierre et Marie Curie, 
   CNRS UMR 7600, F-75252 Paris, France}
\author{J.~Ollivier}
   \affiliation{Institut Laue-Langevin, CS 20156, F-38042 Grenoble Cedex 9, France}
\author{S.~Rols}
   \affiliation{Institut Laue-Langevin, CS 20156, F-38042 Grenoble Cedex 9, France}
\author{P.~Manuel}
   \affiliation{ISIS Facility, Rutherford Appleton Laboratory, Chilton, Didcot, OX11 0QX, United Kingdom}
\author{A.~S.~Wills}
   \affiliation{University College London, Department of Chemistry,  20 Gordon Street, London, WC1H 0AJ, UK}

\date{\today}

\begin{abstract}
The mineral haydeeite, \hayd, is a $S=1/2$ kagome ferromagnet 
that displays long-range magnetic order below $T_C=4.2$~K 
with a strongly reduced moment. 
Our inelastic neutron scattering data  show clear spin-wave excitations 
that are well described by a Heisenberg Hamiltonian 
with ferromagnetic nearest-neighbor exchange $J_1=-38$~K 
and antiferromagnetic exchange $J_d=+11$~K 
across the hexagons of the kagome lattice. 
These values place haydeeite in the ferromagnetic part of the phase diagram, 
close to the non-coplanar twelve-sublattice cuboc2 antiferromagnetic order. 
Diffuse dynamic short-range ferromagnetic correlations observed above $T_C$ 
persist well into the ferromagnetically ordered phase with a behavior distinct from critical scattering. 
\end{abstract}

\maketitle

\begin{figure}[b]
\includegraphics[width=0.99\columnwidth]{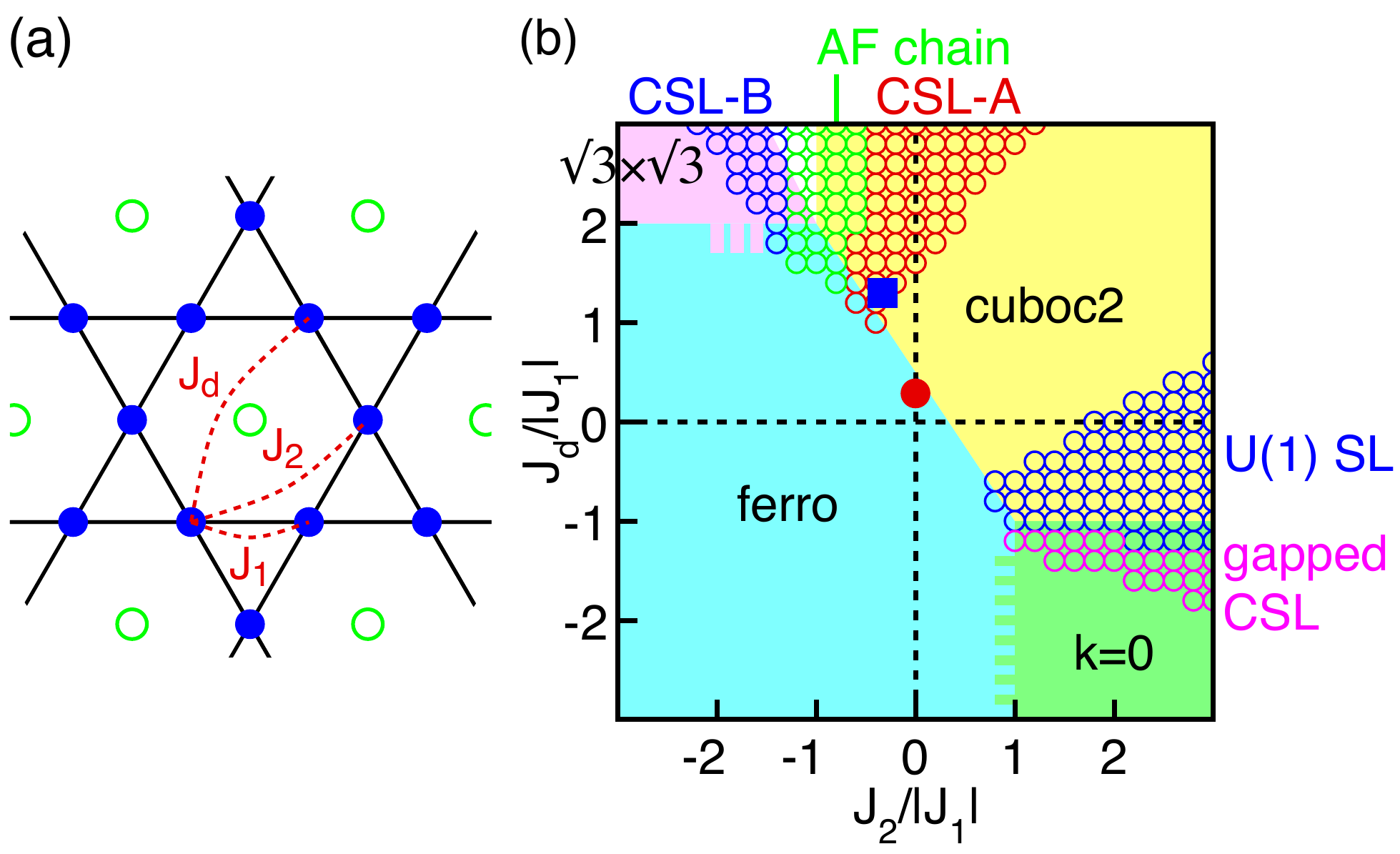}
\caption{(Color online) 
(a) Kagome lattice of haydeeite with Cu$^{2+}$ $S=1/2$ spins (blue solid circles), diamagnetic Mg$^{2+}$ ions (green open circles), 
and main exchange interactions (red dashed lines). 
(b) Phase diagram of the kagome lattice for a ferromagnetic nearest-neighbor interaction, $J_1<0$.
The filled (colored) areas correspond to magnetically ordered phases 
and the open circles show regions where quantum spin liquid states are predicted \cite{Bieri15}. 
The (blue) solid square represents kapellasite \cite{Fak12,Bernu13}
while the (red) solid circle corresponds to haydeeite. }
\label{FigPhaseDia}
\end{figure}

Quantum $S=1/2$ spins localized on the two-dimensional kagome lattice of vertex-sharing triangles 
are highly frustrated magnetic model systems that are predicted to 
form quantum spin liquids (QSLs) with exotic ground states and excitations \cite{Balents10}.
While initial work focused on nearest-neighbor Heisenberg kagome antiferromagnets 
\cite{Hastings00,Ran07,Hermele08,Yan11,Lu11,Iqbal11z2}, 
such as the mineral herbertsmithite \cite{Han12}, 
it has been realized that 
further neighbor interactions or the antisymmetric Dzyaloshinski-Moriya (DM) interaction, 
which is intrinsically allowed in the kagome lattice, may lead to even more fascinating ground states
\cite{Messio12,Jiang12,Messio13,Gong14}. 
Even for the case of classical spins, the kagome lattice with interactions beyond nearest neighbors (NN) [see Fig.\ \ref{FigPhaseDia}(a)] 
shows a very rich magnetic phase diagram with complex non-coplanar structures such as  
chiral and multi-{\bf k} magnetic ordering \cite{Domenge05,Domenge08,Messio11}. 
This is illustrated by the uniformly filled areas in Fig.\ \ref{FigPhaseDia}(b) for the case of ferromagnetic NN interaction $J_1<0$. 
Very recently, it was shown that for quantum spins, QSL phases ``invade'' parts of the classical phase diagram \cite{Suttner14,Gong15,Bieri15}, 
as shown by the open circles in Fig.\ \ref{FigPhaseDia}(b). 

An experimental realization of such a system that exemplifies the importance of quantum effects 
is the mineral kapellasite \cite{Colman08,Colman10,Fak12,Bernu13,Kermarrec14}, 
where high-temperature fits to the magnetic susceptibility \cite{Bernu13} place it in the area of the phase diagram 
that corresponds to the non-coplanar twelve-sublattice magnetically ordered cuboc2 phase [see (blue) solid square in Fig.\ \ref{FigPhaseDia}(b)]. 
However, kapellasite does not actually order magnetically, 
but instead forms a QSL with gapless spinon-like excitations and short-range correlations, 
which are reminiscent of the cuboc2 phase \cite{Fak12}. 
Following recent theoretical work \cite{Bieri15}, 
it can now be understood that kapellasite falls in the region of the phase diagram 
where quantum fluctuations favor a chiral gapless spin liquid state labelled CSL-A, 
which has similar short-range correlations to kapellasite.

In this Rapid Communication, we explore experimentally other parts of the phase diagram shown in Fig.\ \ref{FigPhaseDia}(b). 
By replacing the Zn$^{2+}$ ion [green open circle in Fig.\ \ref{FigPhaseDia}(a)] of kapellasite by Mg$^{2+}$, 
we expect  the $J_d$ exchange integral to change the most. 
Such a substitution leads to the mineral haydeeite, 
which orders ferromagnetically \cite{Colman10}. 
Our inelastic neutron scattering data allow us to characterize the ground state and determine the exchange integrals via spin-wave theory. 
We find that it is the decrease of $J_d/|J_1|$ that places haydeeite in the ferromagnetic part of the phase diagram 
rather than in the cuboc2 antiferromagnetic state. 

Haydeeite, \hay, crystallizes in the trigonal  $P\bar{3}m1$ space group (\#164) with
lattice parameters $a = 6.2885(1)$ and $c = 5.7271(1)$ \AA. 
It is isostructural with kapellasite \cite{Colman08,Colman10} 
and also has $S=1/2$ Cu$^{2+}$ spins forming a highly two-dimensional kagome lattice of equilateral corner-sharing triangles. 
The kagome lattice is sketched in Fig.\ \ref{FigPhaseDia}(a) with the main exchange interactions $J_1$, $J_2$, and $J_d$. 
The diamagnetic Mg$^{2+}$ ions sit at the centers of the hexagons. 
Note that haydeeite should not be confused with its polymorph `Mg-herbertsmithite' 
$\gamma$-MgCu$_3$(OH)$_6$Cl$_2$, 
which has the  same chemical formula but  a different crystallographic structure and different magnetic properties \cite{Kermarrec11}.  

A 1.75~g powder sample of fully deuterated haydeeite was synthesized using 
a scaled-up variant of the established method \cite{Colman10,Boldrin}.  
Powder diffraction using x-rays and neutrons confirms the crystallographic structure \cite{Colman10} and, 
as expected, a higher degree of site order than in kapellasite. 
In haydeeite, there are no Cu ions on the Mg site and about 16\% Mg ions reside on the Cu sites, 
resulting in a slightly diluted kagome lattice.

\begin{figure}[t]
\includegraphics[width=0.99\columnwidth]{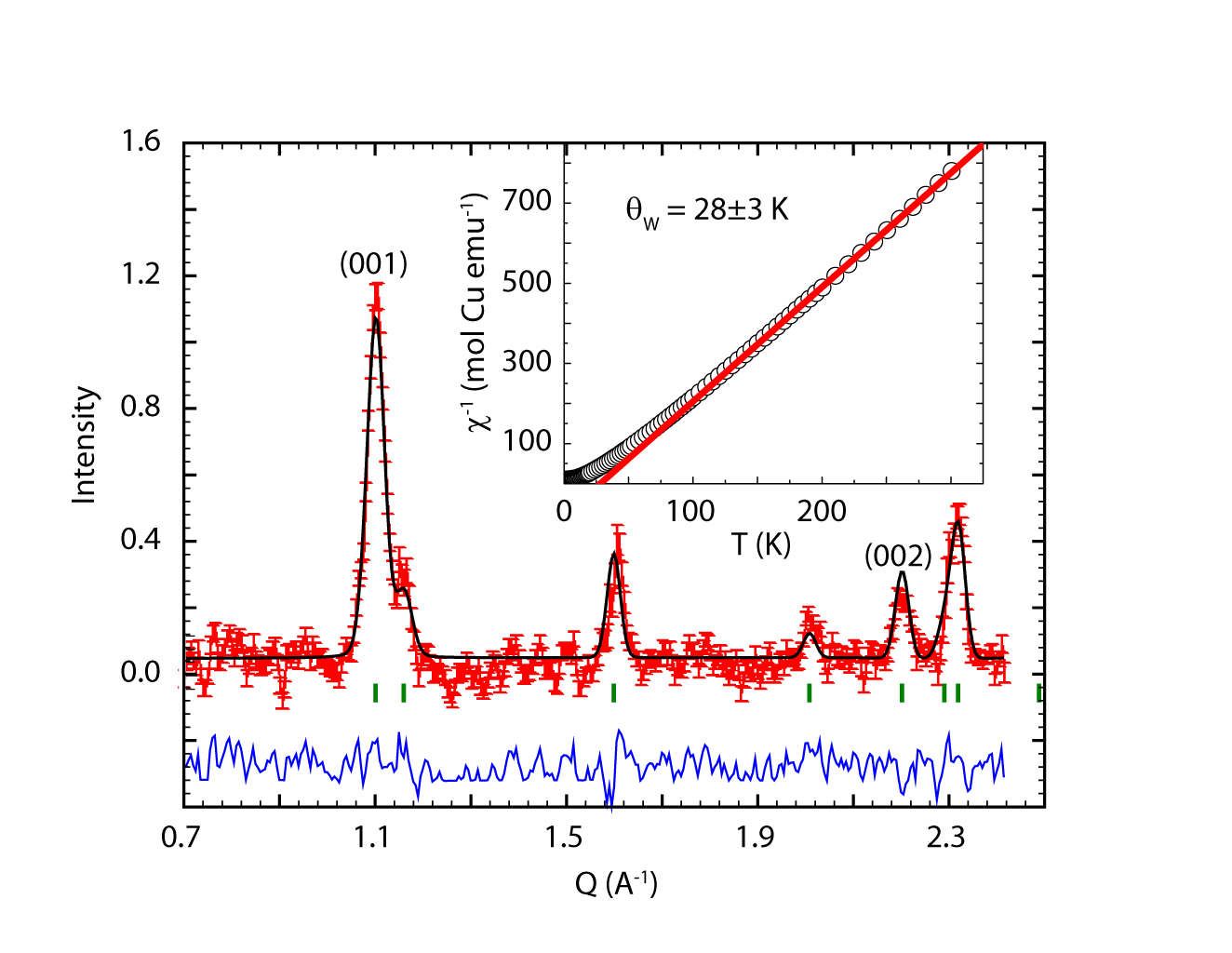}
\caption{(Color online) 
Magnetic Bragg scattering from haydeeite, \hayd, 
obtained by integration over the elastic peak ($|E|<0.02$~meV) of data from IN5 taken at $T=1.5$~K after subtraction of 8~K data. 
The (red) dots are experimental data with one-sigma error bars, 
the solid (black) line going through the data points is the calculated diffraction pattern, 
and the (blue) line at the bottom is the difference. 
The inset shows the inverse susceptibility with a linear fit. }
\label{FigBragg}
\end{figure}

Measurements of the magnetic susceptibility $\chi(T)$ at a field of $H=5$~T and the magnetization $M(H)$ at $T=2$~K 
show that deuterated and hydrogenated haydeeite have very similar magnetic properties: 
Ferromagnetic ordering sets in at $T_C=4.2$~K, 
the saturated moment is 0.83~$\mu_B$ per mole of Cu, 
the Curie-Weiss temperature extracted from a fit of the linear region $150<T<300$~K 
of the inverse susceptibility is $\theta_{\rm CW}=28\pm 3$~K, 
and a drop  in $\chi T$ at low temperatures suggests antiferromagnetic components. 

Inelastic neutron scattering measurements on the fully deuterated \hayd\ powder sample 
were performed on the time-of-flight (TOF) spectrometers IN5 and IN4 at the Institut Laue-Langevin, 
using neutrons with incident energies $E_i$ of 3.55 and 16.6~meV. 
These measurements show both acoustic and optic spin-wave branches in the ferromagnetically 
ordered phase as well as weak magnetic Bragg peaks that are apparent in the difference data (Fig.\ \ref{FigBragg}). 
The high sensitivity, low background, 
and energy selectivity of the inelastic TOF technique allows us to conclude that these Bragg peaks, 
which were not observed by standard (energy-integrating) neutron diffraction techniques, 
probably due to the overlapping strong inelastic diffuse scattering discussed below, 
correspond to a $k=0$ propagation vector. 
Combined with the magnetic susceptibility measurements, this confirms the ferromagnetic ordering. 

However, the inelastic TOF technique is not well suited to actually determine the magnetic structure. 
We have compared on a qualitative level the observed magnetic intensity to that calculated for the three 
irreducible representations (IRs)  \cite{Sarah}. 
The best description is obtained with a collinear magnetic 
structure where the moments are in the kagome planes, 
which corresponds to the two-dimensional  $\Gamma_5$ IR in Kovalev's notation \cite{Kovalev}, see Fig.\ \ref{FigBragg}. 
A rough estimate of the magnetic moment gives a strongly reduced magnitude of 
at most 0.2 $\mu_B$ at $T=1.5$~K, which suggests strong quantum fluctuations.

\begin{figure}[t]
\includegraphics[width=0.99\columnwidth]{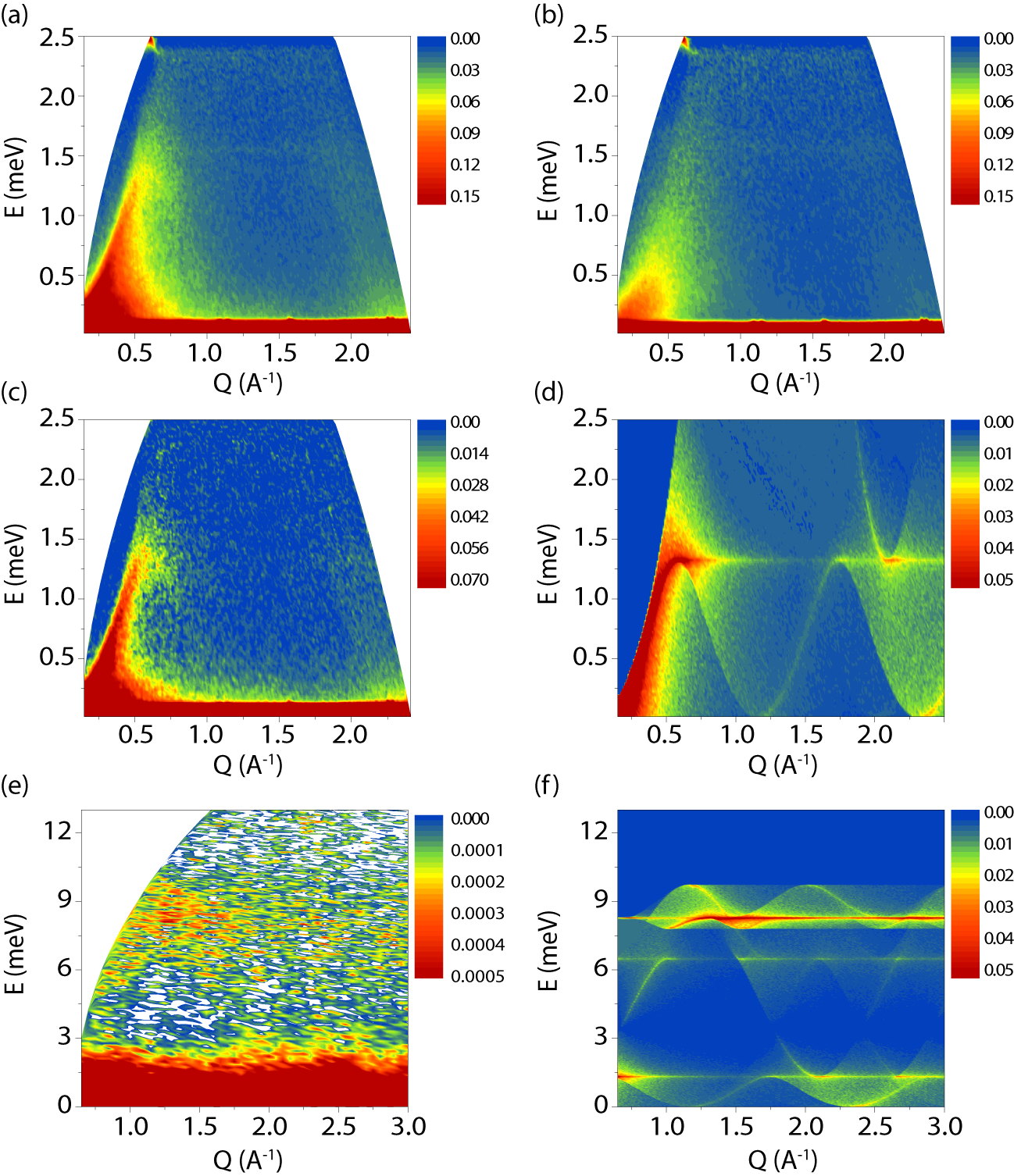}
\caption{(Color online) 
Dynamic susceptibility $\chi^{\prime\prime}(Q,E)$ on a linear intensity scale 
as a function of wave vector $Q$ and energy $E$ of fully deuterated haydeeite.  
Low-energy data measured with $E_i=3.55$~meV are shown in the panels:
(a) ferromagnetic state at $T=1.5$~K, 
(b) paramagnetic state at $T=8$~K, and
(c) magnetic part of the scattering after subtraction of the $T=8$~K data. 
Panel (e) shows the magnetic scattering at higher energies ($E_i=16.6$~meV)
in the ferromagnetic state at $T=2.2$~K 
after subtraction of the high-temperature data at $T=60$~K. 
Spin-wave calculations at low and high energies are shown in panels (d) and (f), respectively.  }
\label{FigMap}
\end{figure}

The observed inelastic neutron scattering intensity is proportional to the dynamic magnetic susceptibility $\chi^{\prime\prime}(Q,E)$, 
shown in Fig.\ \ref{FigMap} for \hayd\ before correction for the magnetic form factor. 
Acoustic spin waves are clearly observed in the low-energy data of the ferromagnetically ordered phase shown in Fig.\ \ref{FigMap}(a), 
and are even more apparent after subtraction of the paramagnetic high-$T$ data of Fig.\ \ref{FigMap}(b), 
see Fig.\ \ref{FigMap}(c). 
Optic spin waves are observed at higher energies, see Fig.\ \ref{FigMap}(e). 

\begin{figure}
\includegraphics[width=0.99\columnwidth]{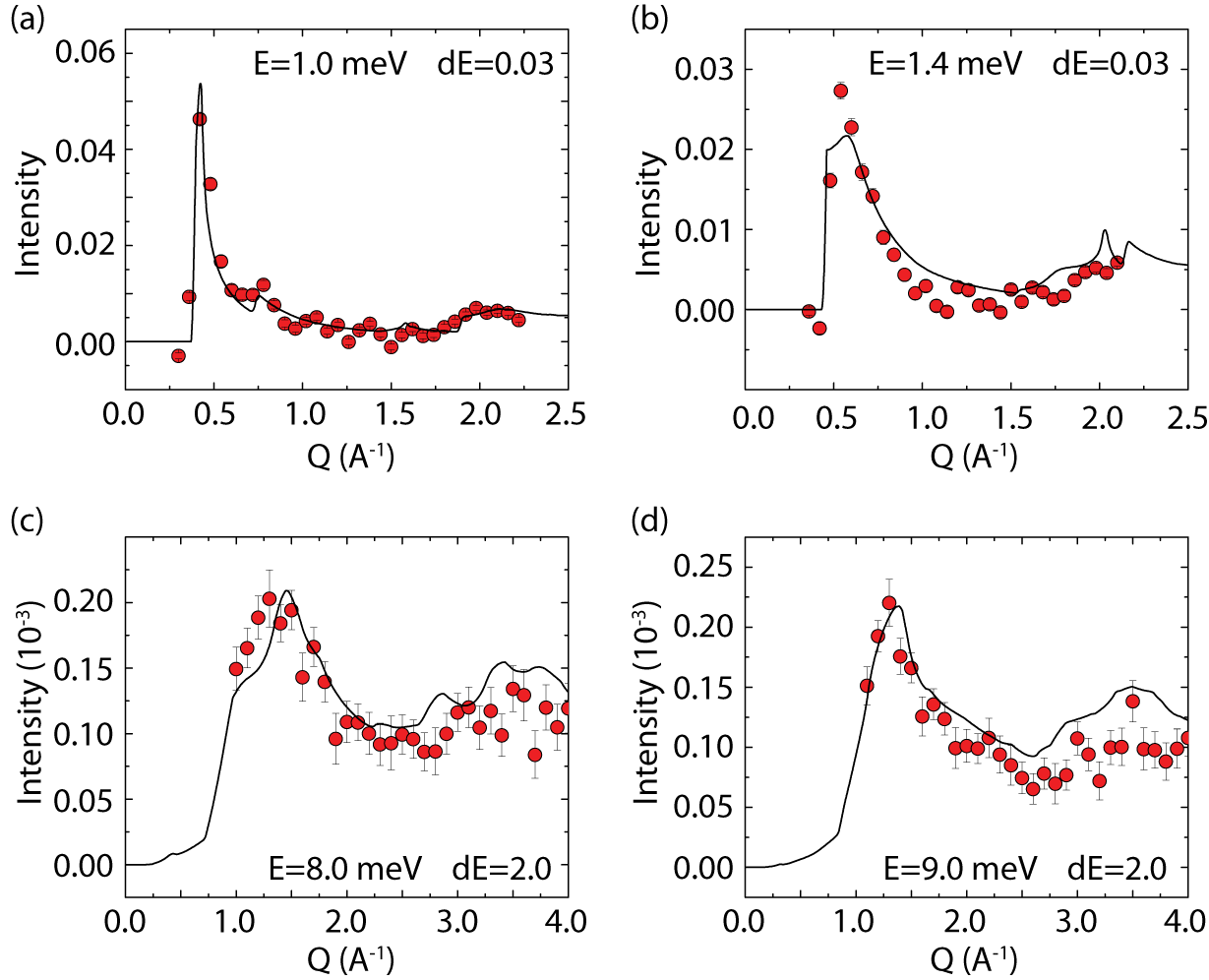}
\caption{(Color online) 
Cuts along $Q$ of the magnetic scattering $\chi^{\prime\prime}(Q,E)$ at constant energies $E$ 
as specified in the ferromagnetically ordered phase of haydeeite \hayd\ (closed circles) 
and the powder-averaged spin-wave calculation for $J_1=-38$ and $J_d=+11$~K (solid line). 
(a)--(b) Excitation energies $E=1.0\pm 0.015$ and $1.4\pm 0.015$~meV measured with $E_i=3.55$~meV at $T=1.5$~K. 
(c)--(d) Excitation energies $E=8\pm 1$ and $9\pm 1$~meV measured with $E_i=16.6$~meV at $T=2.2$~K. 
}
\label{FigQscans}
\end{figure}

The magnetic scattering can be modeled using standard linear spin-wave theory 
for an isotropic Heisenberg Hamiltonian with further-neighbor interactions. 
Since haydeeite has three magnetic ions per primitive magnetic unit cell and a propagation vector $k=0$, 
there are three spin-wave modes, whose energies and intensities are obtained by diagonalizing a real $3\times 3$ matrix. 
Standard Monte Carlo techniques are then used to obtain the powder-averaged dynamic susceptibility  $\chi^{\prime\prime}(Q,E)$. 
The best agreement with the data is found for $J_1=-38$, $J_2=0$, and $J_d=+11$~K, 
where negative exchange integrals correspond to ferromagnetic interactions 
and each bond is counted once in the Hamiltonian. 
The corresponding calculated $\chi^{\prime\prime}(Q,E)$ is shown in Figs.\ \ref{FigMap}(d) and \ref{FigMap}(f). 
The quality of the agreement is clearly seen in $Q$ cuts at constant energy, see Fig.\ \ref{FigQscans}. 
The calculated spectra are very sensitive to the value of $J_2$, 
and from our spin-wave analysis we infer that $|J_2|/|J_1|<<0.1$. 
The Curie-Weiss temperatures calculated from these exchanges is $\theta_{\rm CW}=32$~K, 
in close agreement with the 28~K estimated from the inverse susceptibility. 

Density functional theory (DFT) has predicted a relatively strong antiferromagnetic nearest neighbor interaction $J_1$ 
in both haydeeite and kapellasite \cite{Janson08}, 
in contrast to the ferromagnetic $J_1$ observed experimentally. 
Although the sign of $J_1$ is very sensitive to the O-D distance in the crystallographic structure, 
a ferromagnetic $J_1$ is only achieved in the DFT calculations of haydeeite 
for an O-D distance 20\% shorter than the experimentally determined value of 0.98~\AA\ \cite{Colman10}. 
The origin of this discrepancy is not known. 

\begin{figure}[b]
\includegraphics[width=0.9\columnwidth]{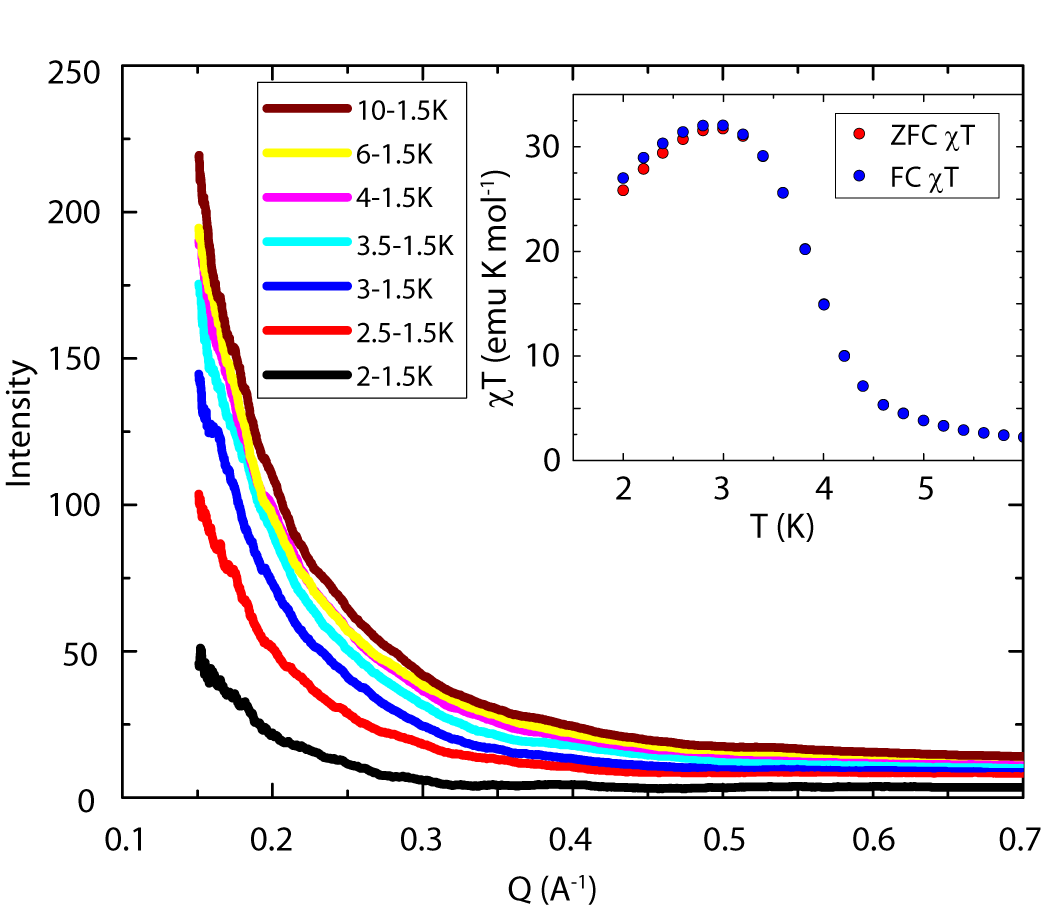}
\caption{(Color online) 
Energy-integrated diffuse scattering from haydeeite for various temperatures after subtraction of the low-$T$ data. 
Inset: Magnetic susceptibility data $\chi T$ {\it vs.}\ $T$ of \hayd\ at $H=5$~T showing ferromagnetic ordering at $T_C=4.2$~K 
and a weak field cooling/zero-field cooling (FC-ZFC) splitting at lower temperatures. }
\label{FigWISH}
\end{figure}

Diffuse inelastic magnetic scattering is clearly observed at temperatures above $T_C$, as shown in Fig.\ \ref{FigMap}(b) at $T=8$~K. 
In contrast to the kagome antiferromagnets kapellasite ($S=1/2$) \cite{Fak12} 
and deuteronium jarosite ($S=5/2$)  \cite{Fak08}, this scattering is clearly dispersive and mimics the spin-wave dispersion, 
thus resembling so-called paramagnons. 
Notably, this diffuse scattering seems to coexist with the spin waves below $T_C$, 
which can be inferred from the anomalously large $Q$-width of the scattering [see Fig.\ \ref{FigMap}(a)]. 
Complementary measurements using the powder diffractometer WISH at ISIS 
show that the diffuse scattering is centered around $Q=0$ and corresponds to short-range ferromagnetic correlations 
with a correlation length of about 35~\AA\ at $T=6$~K (see Fig. \ref{FigWISH}).
The intensity of the short-ranged ferromagnetic correlations is roughly constant between $T=3.5$ and 10~K. 
On further cooling, this intensity decreases, concomitantly with the reduction in $\chi T$  (see inset of Fig.\ \ref{FigWISH}).
The behavior of the rather strong diffuse dynamic magnetic scattering in haydeeite is thus very different 
from what is expected for frozen-in magnetic disorder and also from critical scattering, 
for which the correlation length should diverge at $T_C$. 
The presence of strong short-range correlations close to a quantum phase transition where ferromagnetic order disappears is intriguing.

We will now discuss the role of the Dzyaloshinski-Moriya interaction in haydeeite. 
The DM interaction is of current strong interest in kagome ferromagnets \cite{Katsura10,Matsumoto11,Zhang13,Mook14} 
and kagome antiferromagnets above the saturation field \cite{Starykh},  
since an out-of-plane DM vector opens up gaps at finite energies in the spin-wave spectrum. 
This leads to a topological structure of the magnon bands and a finite Berry  curvature in momentum space, 
resulting in the  magnon Hall effect \cite{Ohgushi00,Onose10}. 
The DM interaction is intrinsic to the kagome lattice, 
since the midpoint of the nearest-neighbor bond lacks inversion symmetry [see Fig.\ \ref{FigPhaseDia}(a)]. 
In haydeeite, just as in the jarosite mineral family \cite{Elhajal02}, 
the CuO$_6$ octahedra of the crystal structure 
are tilted with respect to the kagome planes, 
which allows the DM interaction to have both in-plane, $D_{xy}$, and out-of-plane, $D_z$, components \cite{Elhajal02}. 
We consider the ferromagnetic case with the assumption that the magnitude of the DM interaction is smaller than $|J_1|$, 
and use the experimentally determined exchange integrals (see above). 
For $D_{xy}=0$, the magnetic structure is a collinear ferromagnet. 
A finite value of $D_{xy}$ leads to a $k=0$ umbrella structure with the magnetization perpendicular to the kagome plane 
and where the opening angle of the umbrella increases with increasing $|D_{xy}|$. 
A finite value of $D_z$ creates gaps at finite energies of the spin-wave spectrum between the different branches, 
provided the spins have a component perpendicular to the kagome plane. 
However, our neutron data 
are not compatible with the $\Gamma_3$ IR 
that describes an umbrella-type structure with magnetization perpendicular to the kagome planes. 
Such an umbrella structure should have very low intensity for the (001) and (002) magnetic Bragg peaks, 
whereas the data show strong intensity there (see Fig.\ \ref{FigBragg}). 
Also, no finite-energy gaps are observed in the spin-wave spectrum. 
In other words, there is no experimental evidence that the DM interaction plays a significant role in haydeeite. 

We conclude that haydeeite appears to be a rare example of an {\it isotropic} $S=1/2$ Heisenberg kagome ferromagnet.
The ordered moment is strongly reduced to  $\leq$0.2~$\mu_B$ 
and coexists with diffuse dynamic short-range ferromagnetic correlations below $T_C$. 
We speculate that the reduction of the ordered moment is due to quantum fluctuations in the proximity to the quantum phase transition 
between the Heisenberg kagome ferromagnet and the Heisenberg cuboc2 antiferromagnet. 
Within this picture, a small increase in the third-neighbor across-hexagon antiferromagnetic exchange interaction $J_d$ 
would push haydeeite into the magnetically ordered cuboc2 phase, 
and a further increase of $J_d$ or a slightly ferromagnetic second-neighbor interaction $J_2$ 
would lead to a gapless chiral spin liquid state with spinon Fermi surfaces (CSL-A), 
which breaks reflection and time-reversal symmetries \cite{Bieri15}. 
Such a state may have been observed in kapellasite \cite{Fak12}, 
where the $J_d$ exchange is mediated by Zn$^{2+}$ rather than Mg$^{2+}$ ions.
Studies of these compounds under pressure or with some partial Zn-Mg substitution 
may allow experimental access to quantum spin-1/2 kagome systems at the border of a ferromagnetic instability. 

This work was supported in part by the 
French Agence Nationale de la Recherche Grant No.\ ANR-12-BS04-0021. 
Experiment at the ISIS neutron source was supported by a beam-time allocation 
from the Science and Technology Facilities Council (STFC), UK. 
DB acknowledges financial support from the Institut Laue-Langevin and thanks UCL for provision of a studentship. 
We thank S. Toth, C. Lhuillier, and L. Messio for helpful discussions 
and M.~Zbiri for experimental assistance.

\end{document}